# Structural, Vibrational and Thermodynamic Properties of $Ag_nCu_{34-n}$ Nanoparticles

Handan Yildirim, Abdelkader Kara and Talat. S Rahman
Department of Physics, University of Central Florida, Orlando, FL 32828, USA

## Abstract

We report results of a systematic study of structural, vibrational and thermodynamical properties of 34-atom bimetallic nanoparticles from the $Ag_nCu_{34-n}$ family using model interaction potentials as derived from the embedded atom method and in the harmonic approximation of lattice dynamics. Systematic trends in the bond length and dynamical properties can be explained largely on arguments based on local coordination and elemental environment. Thus increase in the number of silver atoms in a given neighborhood introduces a monotonic increase in bond length while increase of the copper content does the reverse. Moreover, based on bond lengths of the lowest coordinated (6 and 8) copper atoms with their nearest neighbors (Cu atoms), we find that the nanoparticles divide into two groups with average bond length either close to (~ 2.58 Å) or smaller (~ 2.48 Å) than that in bulk copper, accompanied by characteristic features in their vibrational density of states. For the entire set of nanoparticles, vibrational modes are found above the bulk bands of copper/silver. Furthermore, a blue shift in the high frequency end with increasing number of copper atoms in the nanoparticles is traced to a shrinkage of bond lengths from bulk values. The vibrational densities of states at the low frequency end of the spectrum scale linearly with frequency as for single element nanoparticles, however, the effect is more pronounced for these nanoalloys. The Debye temperature was found to be about one third of that of the bulk for pure copper and silver nanoparticles with a non-linear increase with increasing number of copper atoms in the nanoalloys.

## I. Introduction

It has been widely accepted that a solid's surface representing the interface with the surrounding environment introduces differences in properties from that of the bulk form. These differences have opened new avenues through which one can fine-tune the physical and chemical properties and thereby arrive closer to the goal of designing materials with tailored characteristics. Because of their large surface to volume ratio, nanoparticles are good candidates for materials with novel and controllable properties and have been the subject of great interest to the scientific community. Efforts have been made to understand the size-dependent evolution of the physical, chemical and electronic properties of these nanoparticles [1]. These studies also point to the possibility of using nanoparticles as building blocks for cluster assembled materials [2] with tailored properties and possible applications [3] in biomedical, catalytic, optical, and electronic industry [4]. Since alloying offers a natural avenue for further controlling and modifying properties of nanoparticles, attention has been diverted to synthesis, characterization, and observation of novel properties of bimetallic nanoparticles [5]. The latter have also found themselves to be the playground for testing theoretical developments in techniques that aim at sketching multidimensional potential energy surfaces to search for the equilibrium structures of atoms and molecules in complex environments. The competing roles of elemental specificity, relative strengths of bonds and cohesive energy, local coordination, and electronic and geometric structure blend in to provide materials whose stability is not always transparent or easily controllable. Yet, developing an understanding of the microscopic factors that relate structure to functionality is at the base of the increasing important field of computational material design. And nanoalloys provide a bottom-up approach to tailoring the properties of materials through a systematic understanding of the relative importance of the diverse factors that constitute a local environment. The issues at the theoretical level extend from the very determination of the geometric structure of the nanoalloy, to establishment of stability criteria, to extraction of the relative effects of structural, electronic, and vibrational contributions in controlling alloy novel properties. In the case of bimetallic nanoparticles, symmetry and elemental size is also expected to play a role in the determination of geometric structure.

Recently, Rossi et al. [4] applied global optimization techniques to several bimetallic clusters of transition metals elements consisting of 30 to 40 atoms, using semi-empirical inter-atomic potentials [9]. They found very stable "magic" clusters, characterized as core-shell polyicosahedra with high (calculated) melting points. Earlier MD simulations [6] dealing with the growth of the core-shell structures of larger clusters (a few hundreds of atoms) showed that Ni and Cu impurity atoms prefer subsurface locations inside the Ag clusters and induce higher stability and melting temperatures. For the six binary systems (Ag-Ni, Ag-Cu, Au-Cu, Ag-Pd, Ag-Au, and Pd-Pt), Ferrando et *al*, also examine the effects of size mismatch, alloying tendency (as compared to that in the bulk phase like Ag-Ni and Ag-Cu) and the tendency for surface segregation [10,11]. Among these binary systems, Ag-Cu has a large size mismatch between the two atomic species.

These pioneering studies have been very helpful in developing the framework for a systematic methodology for determining the geometric structure and stability criteria for nanoalloys. The stage is now set to supplement the work with complementary studies; in particular, the effect of alloying on the vibrational characteristics of the nanoparticles has yet to be explored. Such an examination is needed because of its relevance in determining relative structural stability as in bulk [7] and surface alloys [8].

Motivated by the above, in this work we have undertaken a systematic examinations of the effect of changing elemental composition on the local structure (bond lengths), vibrational dynamics and thermodynamics for the case of the 34-atom Ag-Cu nanoparticles (the $Ag_nCu_{34-n}$ family). As a starting point we take the geometric structure of each nanoparticle to be that obtained by Rossi et al [4], while recognizing that *a priori* inclusion of vibrational entropy in initial searches for equilibrium configurations could have produced a different outcome. For the rest of the paper, we proceed as follows: In Section 2 we discuss the theoretical details, while Section 3 contains a summary of the results which is divided in to four sub sections. In Section 3.1, we provide the results of the analysis of the bond lengths in the set of nanoparticles, in Section 3.2, we give details of the vibrational densities of states. The analysis of the vibrational free energy, the mean square vibrational amplitudes and Debye temperature are summarized in Sections 3.3 and 3.4, respectively. Finally, in Section 4, we summarize our conclusions.

## 2. Theoretical Details

The starting configurations of the 35 nanoparticles of the family of $Ag_nCu_{34-n}$ were provided by Ferrando *et al* [10] who performed a genetic algorithm scheme to find the minimum energy configuration using an empirical potential [9]. They showed that this family consists of "magic" nanoparticles which are characterized by the common structural property of perfect core-shell with Ag atoms on the surface and Cu atoms inside the shell. In figure 1, we present the structures of six representative nanoparticles showing the differences in the stoichiometry despite similarity in the core shell. In this study, the starting configuration of a given nanoparticle has been subjected to a further relaxation using the conjugate gradient method [12] with interaction potentials derived from the embedded atom method (EAM) [13]. In the case of copper, silver and their alloys, these potentials have proven to provide accurate structural as well as vibrational properties for bulk and surface systems [14]. With the nanoparticles in their equilibrium configuration, we calculate the force constant matrix from the partial second derivatives of the potentials which is necessary to determine the vibrational dynamics and hence the thermodynamical functions. As is well known, the vibrational densities of states may be determined from the force constant matrix (D) for the nanoparticles which allows the extraction of the corresponding Green's function, as shown below. From the trace of the Green's function matrix, the normalized vibrational density of states $\rho(\omega)$ of the nanoparticles is obtained

$$G(\omega^2 + i\varepsilon) = (I(\omega^2 + i\varepsilon) - D)^{-1}$$

$$\rho(\omega^2) = -\frac{1}{\pi}\lim_{\varepsilon \to 0} \mathrm{Im} Tr[G(\omega^2 + i\varepsilon)]$$

$$\rho(\omega) = 2\omega\rho(\omega^2)$$

(1)- (3)

The local densities of states $\rho_i(\omega)$ (densities of states of each atom in the nanoparticles), $\sum_{i=1}^{n} \rho_i(\omega) = \rho(\omega)$, where $n$ is the number of atoms in the nanoparticles, can be determined from the diagonal elements of the imaginary part of Green's function matrix. Once the vibrational densities of states are calculated, we can easily determine the thermodynamical functions in the harmonic approximation of lattice dynamics. The local vibrational free energy for each atom $i$ and their mean square vibrational amplitudes and Debye temperature are thus calculated using the equations below:

$$F_i^{vib} = k_B T \int_0^\infty \ln\left(2\sinh\left(\frac{\hbar\omega}{2k_B T}\right)\right)\rho_i(\omega)d\omega$$

$$\left\langle u_i^2 \right\rangle = \frac{\hbar}{2M} \int_0^\infty \frac{1}{\omega}\coth(\hbar\omega 2k_B T)\rho_i(\omega)d\omega$$

(4) - (5)

$$\theta_i^2 = \frac{3\hbar^2 T}{Mk_B \left\langle u_i^2 \right\rangle}$$

(6)

In consideration of vibrational contributions to the relative stability of the system, the quantity of interest is the excess vibrational free energy which is defined as the excess over the values associated with the bulk system. Thus the local contribution to the vibrational excess free energy is given by:

$$\Delta F_i^{vib} = F_i^{vib} - F_{bulk}^{vib}$$

(7)

where $F_{bulk}^{vib}$ is the bulk value (per atom) for the species (copper and silver) and is obtained from earlier calculations [15]. The total excess free energy (over the bulk), of course, contains a major contribution from the structural or potential energy and from configurational entropy. The excess potential energy part is introduced by Ferrando *et al.* [6] through the term $\Delta$, which is expressed below in Eq. 8. We have similarly added a term which corresponds to the vibrational component of the excess energy ($\Delta_{vib}$).

$$\Delta_{pot} = \frac{E_{tot}^{N,N_1} - N_1\varepsilon_1^{coh} - N_2\varepsilon_2^{coh}}{N^{2/3}} \text{ and } \Delta_{vib} = \frac{F_{tot}^{vib} - N_1 F_{bulk,Ag}^{vib} - N_2 F_{bulk,Cu}^{vib}}{N^{2/3}}$$

(8)

where $E_{tot}^{N,N_1}$ is the minimum energy for a given composition of the nanoparticles, $\varepsilon_1^{coh}$ and $\varepsilon_2^{coh}$ are the cohesive energies of the species (copper and silver), $N_1$ and $N_2$ are the numbers of Ag and Cu atoms and N is the total number of atoms in the nanoparticle. The division by $N^{2/3}$ (approximately the number of surface atoms) yields the excess energy per atom. Here $F_{tot}^{vib}$ is the total vibrational free energy of the nanoparticle:

$$F_{tot}^{vib} = \sum_{i=1}^{N_1} F_{i,Ag}^{vib} + \sum_{i=1}^{N_2} F_{i,Cu}^{vib} \; ; \; F_{bulk,Ag}^{vib} \text{ and } F_{bulk,Cu}^{vib}$$ are the vibrational free energies of the

bulk Ag and Cu atoms, respectively. The total excess energy is now written as:

$$\Delta_{tot} = \Delta_{pot} + \Delta_{vib} \tag{9}$$

On the above we have not included explicitly the contributions from configurational entropy which come into play when deciding the relative stability of isomers corresponding to specific nanoalloy compositions. Such contributions were included by Ferrando *et al,* in their energetic considerations. Our interest here is a specific isomer.

## 3. Results and Discussion

With the objective of investigating the combined effect of alloying, and of the local elemental environment and coordination in determining the structural, vibrational and thermodynamical properties of this set of core-shell nanoparticles, we first provide a global analysis of the bond lengths between atoms in these 35 nanoparticles. This is followed by a presentation of the total vibrational densities of states of the nanoparticles accompanied with local vibrational density of states of chosen atoms from this set of nanoalloys. Global and local analysis of thermodynamical quantities such as free energy, mean square vibrational amplitudes and Debye temperatures are finally presented.

### 3.1. Bond Length Distribution in $Ag_nCu_{34-n}$ Nanoalloys

Since it is neither feasible, nor desirable to present results of the bonding for all atoms in each of the 35 nanoparticles that form the $Ag_nCu_{34-n}$ family, we choose to focus on those with the most dominant coordination. Unlike the case of extended systems, the bond length distribution is very broad and hence a criterion for counting neighbors is necessary. We have set the cut-off in the nearest neighbor bond length count at 2.7 Å and 3.1 Å, for copper and silver atoms, respectively. With these cut offs in mind, we first determine the coordination of all atoms in each nanoparticle. We then determine the total number of Ag and Cu atoms, in the set of 35 nanoparticles with a specific coordination then divide this number by the total number of copper/silver atoms in all nanoparticles put together (34x35) to get the number density. From the plots, in Fig. 2 and 3, of the number density as a function of coordination, we find that copper atoms with coordination 6, 8, 9 and 12 (with their respective number densities of 0.2, 0.15, 0.27 and 0.29) and silver, atoms with coordination 6, 8 and 9 (with number densities of 0.51, 0.23 and 0.12 respectively) are the dominant ones. These number densities reflect the fact that these nanoalloys are core-shell type with Cu atoms populating the core and silver favoring the shell (the highest number densities are for coordination 12 for Cu and 6 for Ag). Moreover, one notes that in these small nanoparticles, only 3% (7%) of the copper (silver) atoms form steps (coordination 7) and that corner/kink atoms with coordination 6 have a large density and even dominate (51%) in the case of silver. As for facets, one notes that for the case of copper atoms, it is the dense (111) geometry that is favorable with 27% of the atoms (coordination 9) versus only 15% for the (100) geometry (atoms with coordination 8). The reverse is true for silver atoms which show a preference for the more open structures with 23% of the atoms on a (100) facet versus only 12% on the (111).

As expected, there is a range of values for the average nearest neighbor bond length for the silver (copper) atoms with the coordination of 6, 8, 9 and (12, 9, 8 and 6) which are discussed in the next paragraph. The change in the average bond lengths for each coordination (changing the number of atoms of either species in the bonding) are shown in Figures 4 and 5 for copper and silver, respectively. From Fig. 4, especially for coordination 12, we find a monotonic increase in the bond length as the number of silver atoms increases in the neighboring bonding. We also found the same behavior for the copper atoms with coordination 9 and 8, while for coordination 6, the statistics are too sparse to draw any specific conclusion.

For copper atoms with coordination 12 and no silver atoms in the neighboring sites, we find the average nearest neighbor bond length to lie between 2.43-2.47 Å, with the exception of the $Cu_{34}$ nanoparticle for which the average bond length is 2.50 Å. With the addition of more Ag atoms to the neighborhood, the average nearest neighbor distance increases monotonically from 2.43 Å (0 Ag neighbor) to 2.70 Å (12 Ag neighbors), corresponding to an increase of 0.02 Å per silver atom. This trend can be explained by the fact that bulk lattice constant of silver is larger than that of copper; consequently, an increase in the number of silver atoms in the neighborhood leads to an increase in their average bond length. This substantial increase will have consequences on the thermodynamical properties of these copper atoms. Note that the spread in the average bond length for any specific environment (for Cu atoms with fixed number of Ag neighbors from any of the nanoparticle) is found to be relatively small (of the order of 0.01 Å to 0.05 Å). For the copper atoms with the coordination 9, the average bond length also increases with the number of silver atoms in the neighboring sites: from 2.54 Å (0 Ag neighbor) to 2.71 Å (6 Ag neighbors). For this coordination, we found the variations in the average bond length for a given neighborhood to be between 0.02 and 0.07 Å. Similarly for copper atoms with coordination 8, we find the total increase in the average bond length is 0.14 Å as the number of neighboring silver atoms changes from 0 (2.53 Å) to 4 (2.67 Å) with a small variation within any given neighborhood (0.03 Å to 0.06 Å). Finally, for coordination 6, the same trend is found as for the case of other coordination: an increase in the bond length with increasing the number of silver atoms. However, since the neighborhood shows a limited range of variation of the number of Ag atoms, the increase in the average bond length is found to be not substantial. For copper atoms with coordination 8 and 6, an interesting trend in bond lengths is found particularly for those no silver atoms in the neighboring sites, for which the bond lengths divide into two groups. For the case of coordination 8, the first group of nanoparticles have the average bond length lying between 2.52 Å and 2.54 Å (smaller than bulk Cu) whereas for the second group, the bond length varies from 2.56 Å to 2.59 Å (larger than bulk Cu). A similar trend (with wider range) is found for coordination 6 for which for the first group of nanoparticles, the bond length lies between 2.48 Å and 2.51 Å. As the number of silver atoms increases in the nanoparticles from 9 to 18 the average bond length varies from 2.52 Å to 2.58 Å with a sudden shift for $Ag_9Cu_{25}$. This is also confirmed by our *ab-initio* electronic structure calculations based on the density functional theory (DFT) which are summarized in Table 1 in parenthesis. We should note that such grouping in the average bond lengths is not found for silver atoms. A more detailed analysis of our results will be presented elsewhere [16] and a DFT study on a particular nanoalloy ($Ag_{27}Cu_7$) is reported in Ref. [17].

As plotted in Fig. 5, the average bond lengths of silver atoms (with change in the number of copper atoms in the bonding) show that the effect of altering the elemental environment is in general much less pronounced than that for copper atoms. We found that there is almost a linear decrease in the bond length with the addition of copper atoms to the bonding. For the case of silver atoms with coordination 9 (7 copper neighbors), appears to violate the above trend (Fig. 5). However, for this coordination, the small number of such atoms in this family of nanoparticles does not produce sufficient statistics to draw a definite conclusion. We also note that there is very little variation in the bond length for silver atoms with no copper atoms in the neighborhood. For the Ag atoms, the average bond length decreases with the addition of copper atoms to the environment from 2.84 Å (0 Cu) to 2.78 Å (5 Cu) for coordination 9, 2.84 Å (0 Cu) to 2.77 Å (6 Cu) for coordination 8 and finally 2.83 Å (0 Cu) to 2.65 Å (6 Cu) for coordination 6.

In summary, for the sets of copper (silver) atoms with coordination 12, 9, 8 and 6 (9, 8 and 6) we find the variation in the bond length to be 0.27, 0.17, 0.14 and 0.1 Å (0.07, 0.09 and 0.2 Å), respectively. We also find characteristic trends in bond length variations induced by alloying, thereby providing the combination of coordination and local elemental environment as its (bond length) measure. We now proceed to examine whether the local coordination plus the environment is an effective measure for determining other characteristics of the nanoalloys.

### 3.2. Vibrational Densities of States of Selected Nanoalloys

When the surface to volume ratio in a solid system becomes high, as is the case of matter at the nano-scale, it is to be expected that a substantial contribution to the properties comes from the surface atoms. Earlier studies on nanoalloys by Ferrando and co-workers [18] showed that the melting temperature of these nanoparticles is relatively high compared to that of single element nanoparticles of the same size. This may be a signature of hardening of the bond between atoms which will have consequences for the vibrational densities of states (VDOS) and hence the thermodynamics of these nanoparticles. It is thus interesting to study the vibrational dynamics and thermodynamics of these systems and correlate any novel features to inter-atomic bonding. It will also be interesting to compare our findings with those already known for Ag and Cu surfaces, nanoparticles and bulk solid [15, 19, 20, 21]

In an earlier study of single element (silver) nanoparticles with varying size (2 to 3.5 nm) [19], two features of the VDOS were found to be different from those of the bulk. The first concerns the high frequency end of the spectrum and the other the low frequency part. It was shown that the high frequency end is shifted above the top of the bulk band and was traced to the shrinking of the nearest neighbor distance of the specific atoms in the nanoparticles [19]. The shift of the high frequency band was shown to be localized and to drop sharply with increasing size of the nanoparticles. At the low frequency end, these nanoparticles exhibited substantial enhancement in the density of states as compared to the bulk spectrum. This enhancement was found to be the contribution of the outer atoms (surface atoms).

In the case of nanoalloys not only the local coordination but also alloying is an important parameter affecting the VDOS. To illustrate the effect of alloying, we present in Fig. 6, the total VDOS of six selected ($Ag_{31}Cu_3$, $Ag_{27}Cu_7$, $Ag_{17}Cu_{17}$, $Ag_{10}Cu_{24}$, $Ag_7Cu_{27}$ and $Ag_3Cu_{31}$) nanoparticles and compare the results to those of single element nanoparticles $Cu_{34}$ and $Ag_{34}$. We note a clear correlation at the high frequency end of the spectrum with the change in the number of copper atoms in the nanoparticles. Starting with Fig. 6a, for the nanoparticles with only three copper atoms, one can note a shift towards the high frequency end, as compared to the case of the pure 34-atom silver nanoparticle. This (alloying) effect becomes stronger for the other cases which have increased alloying of Ag with Cu. One should also note that the shift of the modes towards the high frequency end of the spectrum is found to extend over few THz (over the top of the bulk band). In figures 6e-6f, with increasing number of copper atoms, we note that the spectrum becomes close to that of the single element Cu nanoparticle ($Cu_{34}$), as is to be expected. To quantify the effect of alloying in these nanoparticles, we have calculated the percentage (relative partial integral) of the VDOS in the high frequency region above 5.1 THz, the frequency at which the VDOS of $Ag_{34}$ is almost zero. In figure 7, we show this percentage as a function of the number of copper atoms in the system and find the dependence to be almost linear. The small deviation from linear behavior in Fig. 7 is understandable given the simplicity with which the information has been extracted. The behavior at the low frequency end of the spectrum for the chosen nanoalloys is found to be similar to that for single element nanoparticles for which the VDOS scales linearly with the frequency [19].

Below, we focus on further examination of the effect of environment on the dynamical properties of the nanoalloys by calculating the local vibrational densities of states (LVDOS) for chosen copper (silver) atoms with coordination 12, 9, 8 and 6 (9, 8 and 6) respectively.

### 3.2.1. Local Vibrational Density of States of Copper Atoms in Selected Nanoparticles

The analysis of the local vibrational dynamics relies on the analysis of the bond length presented. The vibrational dynamics of copper atoms, from nanoparticles of selected composition, with coordination 12, 9, 8 and 6, with varying environment, along with the bulk densities of states (of copper atom) are shown in figures 8a-8e. For atoms with coordinations other than 6, we have picked at random one such atom from the family of the nanoparticles, since we find very little spread in the bond lengths. However, for coordination 6, we need to be careful since the bond lengths (see Table1) fall in two categories "larger than bulk value" and "smaller than bulk value", as discussed above. We examine the vibrational dynamics of representative atoms from each of the two categories to ascertain any differences in the spectra.

From the LVDOS of copper atoms with coordination 12 (in figure 8a), the first observation we make is the appearance of new peaks at 8.5 and 9.1 THz for environments containing 0 and 6 silver neighbors (no peaks appear in this high frequencies range if there are 12 silver neighbors). The latter reflects softening in the Cu-Ag coupling introduced by the larger number of silver atoms. At the low frequency end, we note an enhancement of the density of states, reflecting the induced softening of the bond by the presence of silver atoms. These features can be rationalized using bond length arguments. We note that the bond length for copper atom in these three environments is in the range of (2.41-2.53 Å) for 0, (2.50-2.74 Å) for 6 and finally (2.64-2.74 Å) for 12 silver neighbors. The shorter bond lengths are responsible for the high frequency peaks at 8.5 and 9.1 THz (0 and 6 silver neighbors). For 12 silver neighbors, the shortest bond length (2.64 Å) is higher than the bond length in bulk copper (2.56 Å) thus explaining the absence of modes with frequencies above the bulk Cu band in Fig. 8a. Larger bond lengths tend to reveal a softer bonding which enhances the low frequency end of the density of states. Hence, for those copper atoms with 6 and 12 silver neighbors with large bond length (2.74 Å), there is an enhancement around 3 THz, which is absent for 0 silver neighbor whose maximum bond length is 2.53 Å. Similar arguments explain the features in the LVDOS of copper atoms with coordination 9, with (0, 3 and 6 silver neighbors) in figure 8b.

Let us now turn to the case of copper atoms with coordination 8 (having 0, 2 and 4 silver neighbors) for which the LVDOS is shown in figure 8c. At the high frequency end, the largest shift is found for the case of Cu atom with 2 silver neighbors followed by those with 4 silver neighbors. The smallest shift in the high frequency region is found for Cu atom with 0 silver neighbors (bond lengths range 2.40 to 2.80 Å, with 2 Ag neighbors 2.41 to 2.78 Å and with 4 silver neighbors 2.43 to 2.82 Å**.** These results do not follow the trend seen above. The effect of local environment on the features obtained in the high frequency end of the LVDOS thus point to the need for more accurate analysis based on electronic structure calculations. At the low frequency end, however, there is more pronounced deviation of the VDOS from the bulk with loss of coordination: the increase of silver atoms enhances the shift towards low frequency.

Finally, we discuss the vibrational dynamics of atoms with coordination 6. In figure 8d, the LVDOS is shown (for the first category of large bond length) with 0, 1 and 2 silver neighbors in the environment. At the high frequency end, the largest shift is found for atoms with 0 and 1 silver neighbor (the shortest bond lengths are 2.29 and 2.31Å respectively). For Cu atoms with 2 silver neighbors, the shortest bond length is 2.41 Å which has the smallest shift at the high frequency end. For the low frequency end, one notes the complexity of the spectra. However, we can extract a few trends: the order of enhancement is for Cu atoms with 0, 1, and 2 silver neighbors (the largest bond length being 2.65 Å, 2.75 Å, 2.73 Å), respectively.

One would expect intuitively that the largest bond-length would cause the largest softening, however; this is not the case here. In figure 8e, we report the LVDOS of atoms with coordination 6 (from the category of short bond lengths) containing 0, 1 and 2 silver neighbors. We note that a shift in high frequencies is found only for copper atoms with 2 silver neighbors (Table 1). It is interesting to note that even for the atoms with 0 and 1 silver neighbor with relatively short bond lengths (2.44 and 2.39Å respectively) the corresponding LVDOS does not show any new features at high frequencies. For the low frequency end, we note the same trend (as it is shown in figure 8d) for which increasing the number of silver neighbors combined with the highest bond length, such as, for 2 silver (2.73 Å), 1 silver (2.64 Å) and 0 silver neighbors (2.52 Å) does not yield enhancement of low frequency modes. In summary, while useful trends are obtained, the above analysis also points to the fact that complex correlations between coordination and environment introduce some features in the local vibrational densities of states that can not be rationalized on the basis of local environment and bond-length arguments. Further work is needed to address these finer points.

### 3.2.2. Local Vibrational Density of States of Silver Atoms in Selected Nanoparticles

In figures 9a-9c, we present the LVDOS of silver atoms with coordination 9, 8 and 6, following the same procedure used for copper. Here again the choice of the nanoparticle for the atom in question is reported in the figures. For example, for coordination 9 the Ag atom with 0 copper neighbors is chosen from $Ag_{34}$, while those with 3 and 5 copper neighbors are from $Ag_{23}Cu_{11}$. From figure 9a, we see a shift of the density of states towards frequencies above the bulk Cu band for Ag atoms with 3 and 5 copper neighbors, the higher shift being for the latter which can be explained using the bond length argument as above. On the other hand, the enhancement at the low frequency end simply reflects the reduction in the coordination. From figure 9b, we conclude that the features at high and low frequencies of atoms with coordination 8 are quite similar to those with coordination 9.

The most striking features at both high and low frequency ends of the spectrum are associated with silver atoms with coordination 6. As for coordination 9 and 8, the high frequency end contains new modes that are well resolved and arise from environments in which copper is present (figure 9c). The lowest bond lengths are 2.47, 2.5 and 2.77 Å for 6, 3 and 0 copper neighbors, respectively. Note that the bond length associated with 0 copper neighbors is distinct and larger than that with 3 and 6 copper neighbors which, in turn explains the presence of high frequency modes for the last two cases. It is at the low frequency end that we notice a large enhancement for all three cases that is much more pronounced than that of coordination 9 and 8. In addition, however, from the same arguments used for coordination 9 and 8, one would expect to find enhancement for atoms with the highest number of silver neighbors. We note that this is not the case and instead, we see that the highest enhancement of the density of states at low frequency end of the spectrum is associated with a silver atom with 6 copper neighbors.

### 3.3. Vibrational Contributions to the Excess Free Energy

The stability of a system and its phase transitions may be explored by studying its thermodynamical properties in which the vibrational dynamics of a system may play a substantial role. For instance, the role of vibrational free energy in searching for the equilibrium structures of bulk alloy systems has been explored and found to be important [22]. A more recent study of stepped metal surfaces of Cu and Ag showed the vibrational contributions to be a substantial fraction of the step free energy [20]. An equally important and sometimes dominant contribution may arise from configurational entropy. If we were to be comparing the relative stability of a set of isomers for each nanoparticle of a given composition, we would need to include a discussion of the role of configurational entropy. In this work, we are restricting ourselves to the specific nanoparticle configuration that Ferrrando *et al* [10] have already deemed to the candidates of lowest energy for a given set of isomers. We will thus focus only on the assessment of vibrational entropy of a nanoparticle of a fixed configuration.

Our task here is to consider whether the above trend in the contribution of the local vibrational thermodynamic properties to the excess free energy applies also to nanoparticles with only 34 atoms and whether alloying brings some additional characteristics. For this purpose, we have calculated the contribution to the vibrational free energy from each nanoparticle (Table 2) and find that it increases monotonically with the number of copper atoms in the nanoparticle. Since contributions to the low frequency end of the spectrum in the VDOS determine mostly the vibrational free energy at (relatively) low temperatures we have calculated the percentage contribution of this low frequency end to the whole spectrum. To do this, we have used the Debye frequencies of bulk copper and silver (6.56 and 4.48 THz, respectively) as cut-off for the low frequency "part" of the LVDOS and their percentage contributions are summarized in Table 2. It is seen clearly from this table that increasing the number of copper atoms (alloying) depletes the low frequency end of the spectrum resulting in an increase of the contribution of the vibrational dynamics to the free energy. Note that using Debye frequency of bulk copper gives the same trend with higher contributions.

In Ref. [20] it is found that the lower the coordination of the atom (on stepped Cu and Ag surfaces), the higher is its contribution to the excess free energy. It is also reported that the atom below the step (BNN) has lower contribution than that of the atoms in the bulk (one would expect bulk atoms to have the lowest contribution) resulting from over-coordination induced by a strong structural relaxations. To be able to determine how coordination affects the excess free energy, we focus on copper atoms in $Cu_{34}$ nanoparticle (0Ag neighbors). For the coordination studied here (12, 9, 8 and 6) the corresponding local excess vibrational free energy was found to be +12, -6, -6, and -15 meV per atom, respectively. The trend in the contribution of the under coordinated atoms is thus in accord with what has been reported earlier [20]. The most interesting contribution comes from coordination 12 whose opposite sign reflects the effect of over coordination, again in agreement with the earlier study on extended systems except that here the contribution is much larger. For nanoparticles, the over-coordination of the inner atoms comes from the global shrinking that the finite size systems experience [19]. For the case of silver atoms with coordination 9, 8 and 6 we turn to the contribution of the atoms in $Ag_{34}$ nanoparticle (0 Cu neighbors). We find the contribution to be 0, -4 and -7 meV/atom for coordination 9, 8 and 6, respectively. The trend observed here is qualitatively similar to that found for the metal surfaces [20]. We summarize our results for the vibrational entropic contribution for the full set of 35 nanoparticles in Fig. 10 which shows that the inclusion of vibrational contribution to the free energy does not introduce noticeable changes in the relative quantities for these nanoparticles, with optimized geometry. Clearly, the complexity in the local environments leads to both positive and negative contributions to vibrational entropy, making the total contribution small relative to that of the structural energy of the nanoparticle.

### 3.4. Mean Square Vibrational Amplitudes and Debye Temperature

The vibrational densities of states of the studied set of nanoparticle family showed dramatic changes with increasing alloying. In this section we examine the mean square vibrational amplitudes and the Debye temperature of selected atoms in the nanoalloy. In an earlier study on vicinal metal surfaces [21] the authors showed the effect of coordination on mean square vibrational amplitudes and the Debye temperature for atoms with a variety of coordination and concluded that the mean square vibrational amplitudes is enhanced for low coordinated atoms and that Debye temperatures is reduced to about 2/3 of the bulk value.

The local mean square vibrational amplitudes calculated within the harmonic approximation for copper (silver) atoms with coordination 12, 9, 8 and 6 (9, 8 and 6) are presented in figures 11(12). For the sake of simplicity, we will only discuss deviations from the bulk value at 300 K. Starting with copper atoms, for coordination 12, see Fig. 11a, the largest deviation (0.005 $Å^2$) from the bulk value (0.0325 $Å^2$) is found for the case which has the highest number of silver nearest neighbors. For coordination 9 and 8, the more the number of silver neighbors the higher does the mean square vibrational amplitudes deviates from the bulk (0.022 $Å^2$ and 0.04 $Å^2$ for coordination 9 and 8, respectively). The for which a trend opposite to that of atoms with other coordination is seen, as increasing the number of silver atoms results in a decrease in the mean square vibrational amplitudes. This trend can be traced back to the low frequency part of the VDOS (Fig. 8d) which resemble the bulk VDOS. The largest deviation is found to be 0.15 $Å^2$ which is 30, 7 and 4 times larger than that of coordination 12, 9 and 8, respectively. For the short bond length case, the same trend is observed and reported in figure 11e, with a deviation of 0.1 $Å^2$ (for 0 Ag neighbors) which is 20, 5, and 2.5 times larger than that of coordination 12, 9 and 8. Turning to the case of silver atoms, the mean square vibrational amplitudes is shown in Figures 12a-12c for coordination 9, 8 and 6. From figure 12a, we note that a decrease in the number of copper neighbors causes an increase in the mean square vibrational amplitudes with a deviation from bulk of about 0.04 $Å^2$. When the coordination is reduced to 8 (figure 12b), regardless of the number of copper neighbors, the mean square amplitudes become almost the same (0.075 $Å^2$ for 2, 0.073 $Å^2$, 0 and 4 copper neighbors) and the deviation is found to be the same as coordination 9 (0.04 $Å^2$). We thus conclude that local coordination has less effect on the mean square vibrational amplitudes, of silver than for copper atoms, as already reported [20]. For coordination 6, from figure 12c, as in the case of copper atom for the same coordination, we see that addition of more copper atoms induces larger mean square vibrational amplitudes with a deviation of 0.1 $Å^2$.

Note that the mean square vibrational amplitudes being a local quantity, changes dramatically from one atom to another. Experimentally, however, it may not be possible to measure individual mean square vibrational amplitudes but rather a global quantity reflecting an average quantity. In the harmonic approximation, the mean square vibrational amplitudes of any atom scales linearly with temperature and its slope is associated with a Debye temperature as shown in Eq. 6. We have calculated the Debye temperature of each nanoparticle studied here using the mean square vibrational amplitudes of all the atoms for the nanoparticle. This *average* Debye temperature is reported in Table 3 along with a break up into contributions from copper and silver atoms. The upper and lower limits of the average Debye temperature are found to be 69 and 88K, respectively. For copper (silver) atoms, the average Debye temperatures are found to be in the range of 86-111K (60-71K) which is about one third of the Debye temperature of the corresponding bulk. From Table 3, it is clear that the average Debye Temperature of the nanoparticles does increase with the increase of copper atoms; however, increase does not show a linear behavior.

In earlier calculations of the vibrational and thermodynamical properties of the vicinal surfaces [21], the critical role of the coordination in controlling local dynamics of these surfaces was pointed out and rationalized, specifically for coordination 7 to 11 for which a Debye temperature about 2/3 that of the bulk was reported as well as a trend of an increase of the Debye temperature with coordination. Following the same idea, the Debye temperatures for copper and silver atoms with different coordination from 6 to 12 are calculated for the systems of interest here. For clarity again, we discuss here only the Debye temperatures of atoms from $Cu_{34}$ (0Ag) and $Ag_{34}$ (0Cu) nanoparticles. We find for copper atoms with coordination 12, 9, 8 and 6 a Debye temperature 103, 83, 81-96, 73-87 K, respectively. The Debye temperature for coordination 8 and 6 have upper and lower limit ranges reflecting the complicated bonding of this specific coordination as has already been discussed. If we consider the lower limit in the temperature range, we see the same correlations as observed for the vicinal surfaces [21]. For silver atoms with the same coordination, we have found the Debye temperature to be 87, 75, 72 and 62 K for coordination 12, 9, 8 and 6, respectively. The correlation between the Debye temperature and coordination found for copper is also observed for silver atoms.

## 4. Conclusions

A detailed study of the family of $Ag_nCu_{34-n}$ nanoparticles showed novel and rich trends upon alloying with the bonding, vibrational dynamics and thermodynamics (vibrational free energy, local vibrational mean square amplitudes and Debye temperatures) as the main properties explored. For this study, we have taken in to account the coordination and the elemental environment to explain and rationalize the properties of these nano-alloys.

In particular, we find that the average bond length for copper atoms to depend strongly on both coordination and elemental environment. The average bond length monotonically increases by 0.25 Å for environments containing between 0 and 12 Ag neighbors. However, these variations in the average bond lengths were found to be less pronounced for silver atoms. For low coordinated copper atoms (6 and 8), a global analysis of the bond length reveals two regions (short and large average bond length). This bi-modal behavior which goes beyond coordination and direct environment is actually the subject of a more thorough analysis which involves a second degree environmental analysis (second neighbors) and will be reported elsewhere along with a detailed study using density functional theory [16].

We found that increasing the ratio of the number of copper atoms to the number of silver atoms (alloying) induced systematic stiffening in the force field yielding a shift towards high frequencies in the vibrational density of states. On the other hand, we have found the low frequency end of the spectrum to be similar to those of single element nanoparticles which show a linear dependency on the frequency.

We have examined the effect of alloying on the (total) vibrational free energy of the family of $Ag_nCu_{34-n}$ nanoparticles and found it to increase monotonically with the number of copper atoms in the system. The effect of coordination and elemental environment on the excess vibrational free energy showed qualitative similarities with those of atoms on vicinal surfaces and in single element nanoparticles, though substantial quantitative differences appear. The local vibrational mean square amplitudes present strong correlations with coordination. We also found that the calculated average Debye temperature for these nanoparticles changes (increases) with the number of copper atoms in the nanoparticle, however, the change is not linear as it was found for the vibrational free energies. For these nanoparticles, the Debye temperature of copper (silver) atoms in the nanoparticles was found to be about one third of the bulk, as compared to a ratio of two third found for atoms on vicinal surfaces.

## Acknowledgement

We thank Riccardo Ferrando for providing us the initial configuration of the nanoalloys and to Sophie Exdell for help in visualizing and analyzing the data set and resulting multitude of diagrams. T. S. R appreciates fruitful discussions with Miguel Kiwi. This work was supported in part by DOE under grant DE-FG02-07ER46354.

**<u>Figure Captions</u>**

**Figure1.** Structure of some of the chosen nanoparticles a) $Ag_9Cu_{25}$ b) $Ag_{12}Cu_{22}$ c) $Ag_{16}Cu_{18}$ d) $Ag_{17}Cu_{17}$ e) $Ag_{31}Cu_3$  f) $Ag_{33}Cu_1$

**Figure2.** Number density for copper atoms of different coordination

**Figure3.** Number density for silver atoms of different coordination

**Figure4.** Variation of average nearest neighbor bond length of Cu atoms with different coordination and elemental environment

**Figure5.** Variation of average nearest neighbor bond length of Ag atoms with different coordination and elemental environment

**Figure6.** Vibrational densities of states (VDOS) of the chosen nanoparticles
**a)** $Ag_{31}Cu_3$, **b)** $Ag_{27}Cu_7$, **c)** $Ag_{17}Cu_{17}$, **d)** $Ag_{10}Cu_{24}$, **e)** $Ag_7Cu_{27}$, **f)** $Ag_3Cu_{31}$

**Figure7.** Percentage shift in the high frequency end (above 5.1 THZ) for chosen nanoparticles

**Figure8.** Local Vibrational Densities of States (LVDOS) of chosen copper atoms with different coordination **a)**12,  **b)**9, **c)**8, and **d)**6 (Large bond length) **e)**6 (Short bond length) and elemental environment

**Figure9.** Local Vibrational Densities of States (LVDOS) of chosen silver atoms with different coordination **a)**9, **b)**8, and **c)**6 and elemental environment

**Figure10.** The potential, vibrational and total excess energy of the nanoparticles

**Figure11.** Mean Square Displacements of copper atoms with different coordination **a)**12, **b)**9, **c)**8,  **d)**6(Large bond length) and **e)**6 (Short bond length) and elemental environment

**Figure12.** Mean Square Displacements of silver atoms with different coordination **a)**9, **b)**8 and **c)**6 and elemental environment

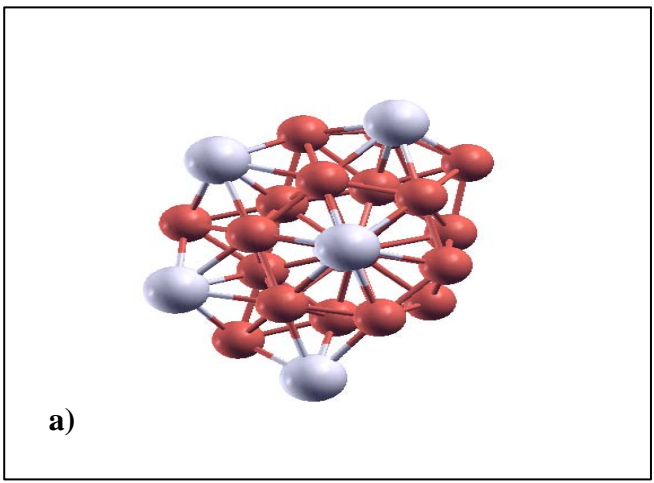

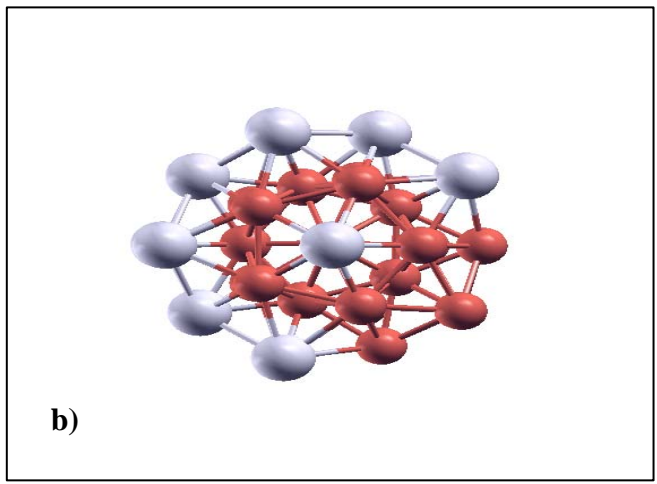

a)                          b)

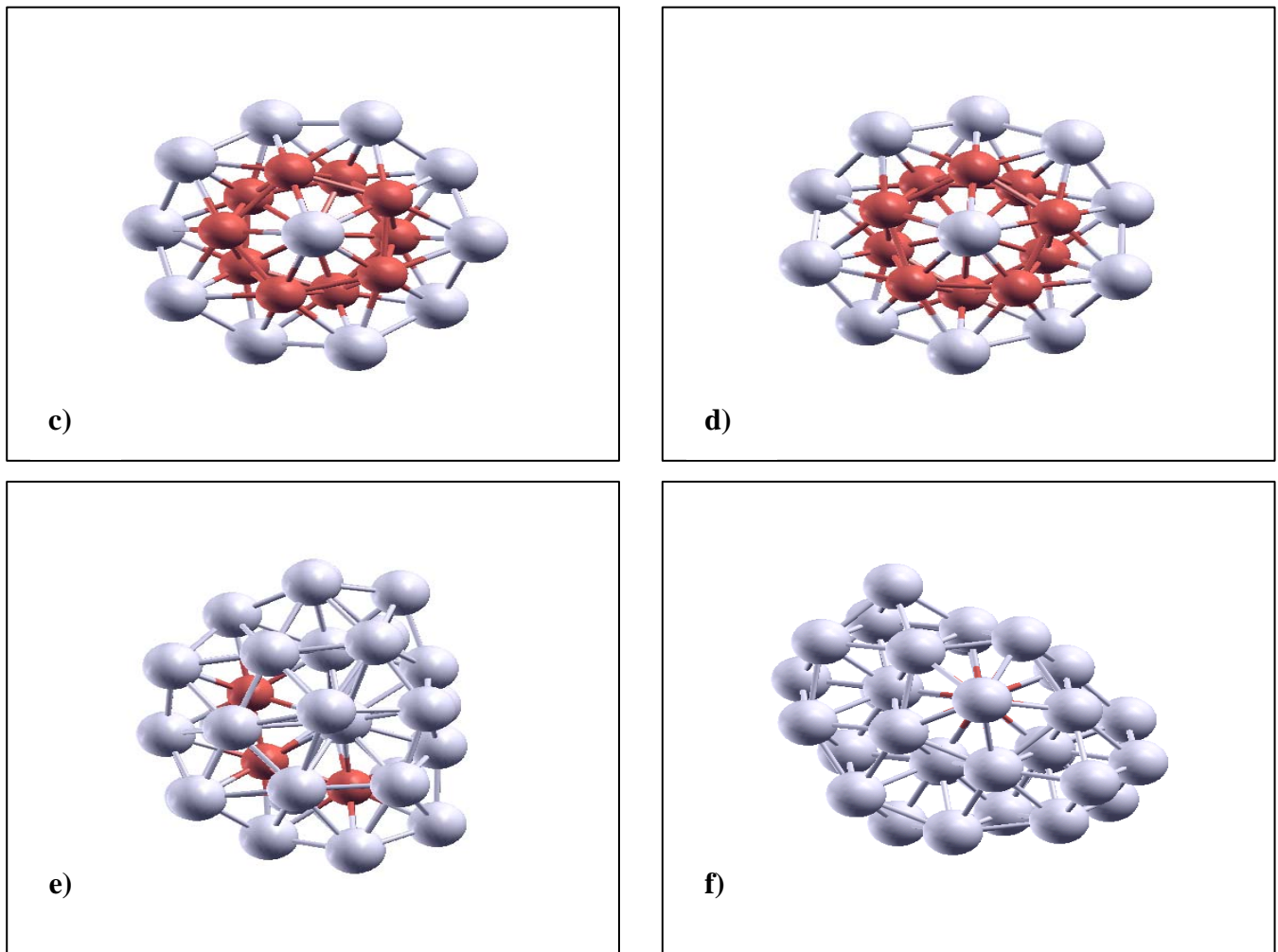

**Figure1.** Structure of some of the chosen nanoparticles
**a)** $Ag_9Cu_{25}$  **b)** $Ag_{12}Cu_{22}$ **c)** $Ag_{16}Cu_{18}$ **d)** $Ag_{17}Cu_{17}$  **e)** $Ag_{31}Cu_3$  **f)** $Ag_{33}Cu_1$

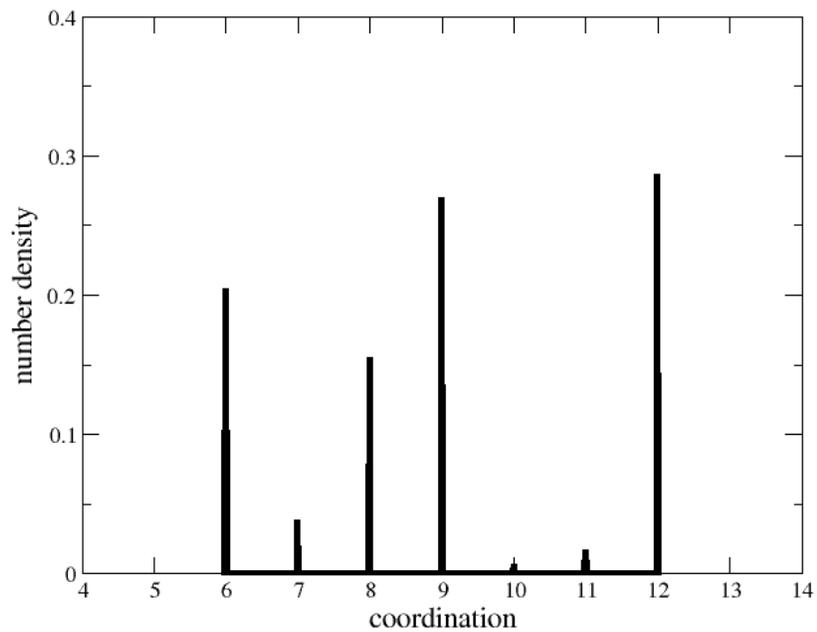

**Figure2.** Number density for copper atoms with different coordination

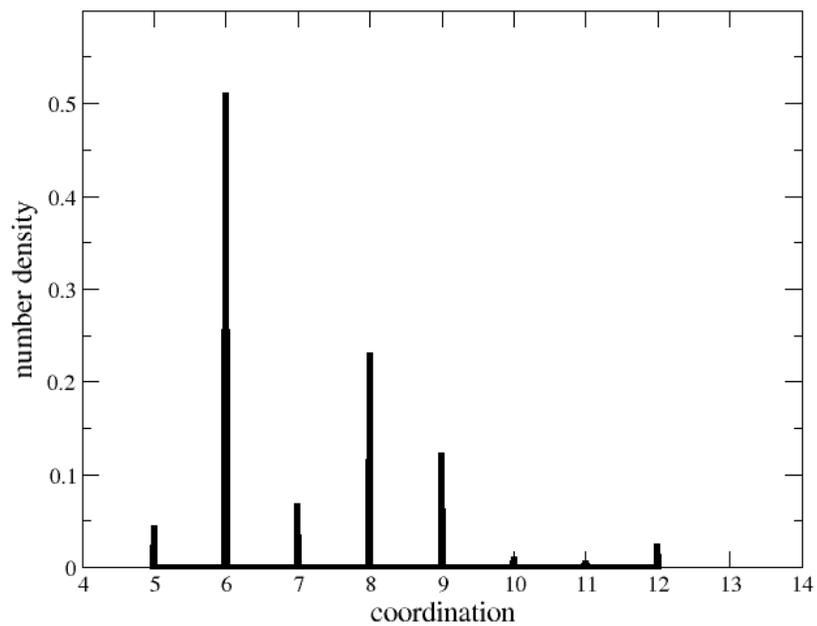

**Figure3.** Number density for silver atoms with different coordination

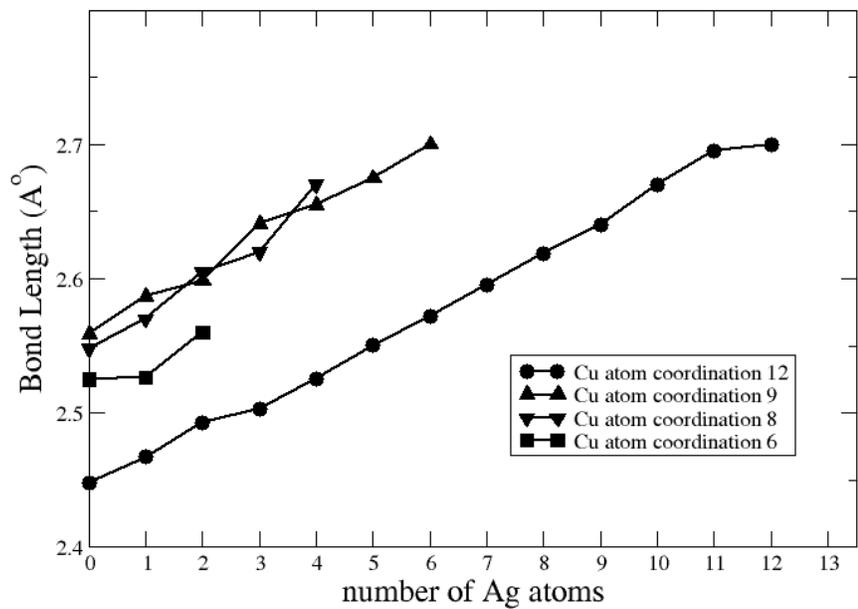

**Figure4.** Variation of average nearest neighbor bond length of Cu atoms with different coordination and elemental environment

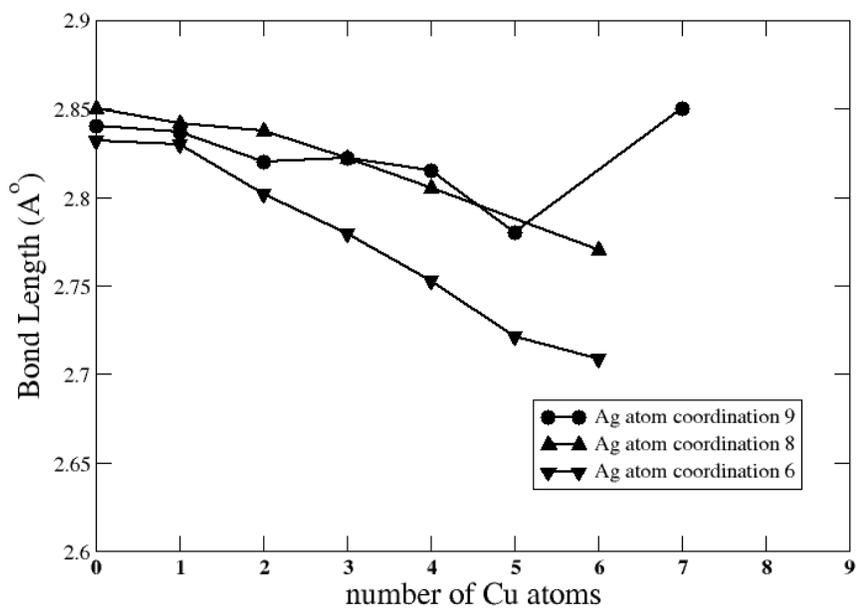

**Figure5.** Variation of average nearest neighbor bond length of Ag atoms with different coordination and elemental environment

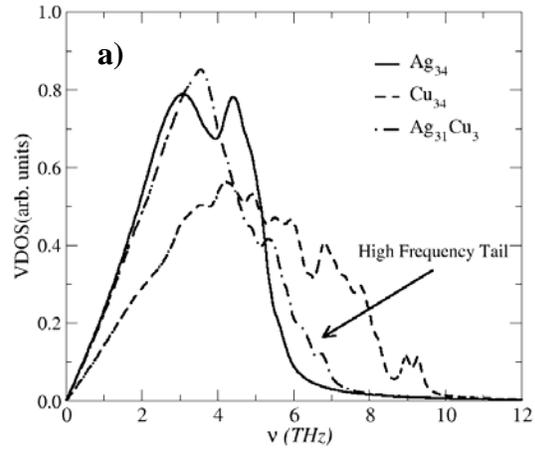

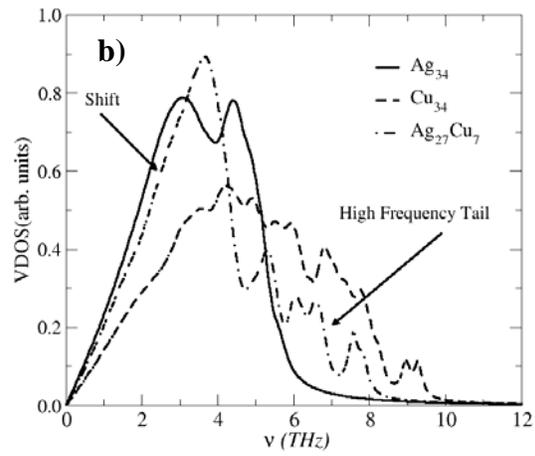

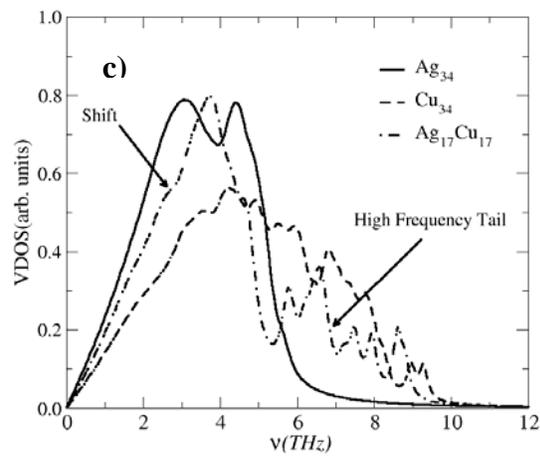

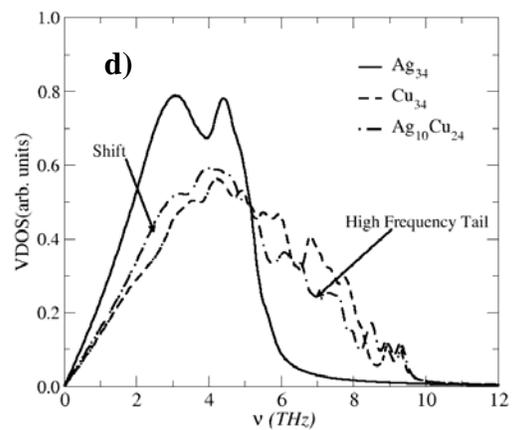

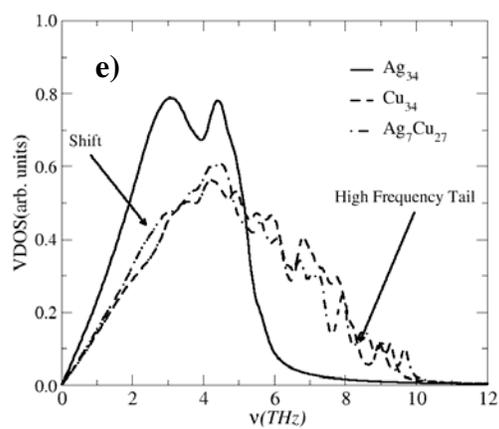

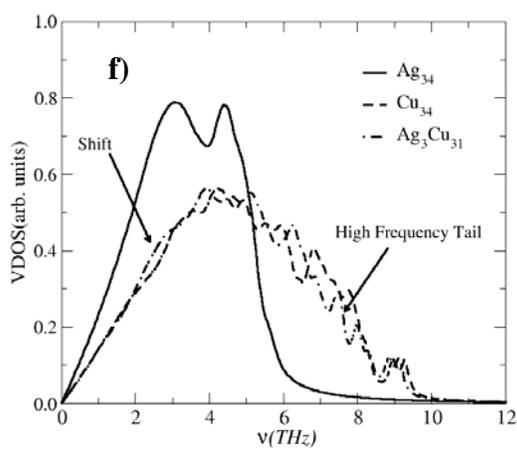

**Figure6.** Vibrational densities of states (VDOS) of the chosen nanoparticles
**a)** Ag$_{31}$Cu$_{3,}$ **b)** Ag$_{27}$Cu$_7$, **c)** Ag$_{17}$Cu$_{17}$, **d)** Ag$_{10}$Cu$_{24}$, **e)** Ag$_7$Cu$_{27}$, **f)** Ag$_3$Cu$_{31}$

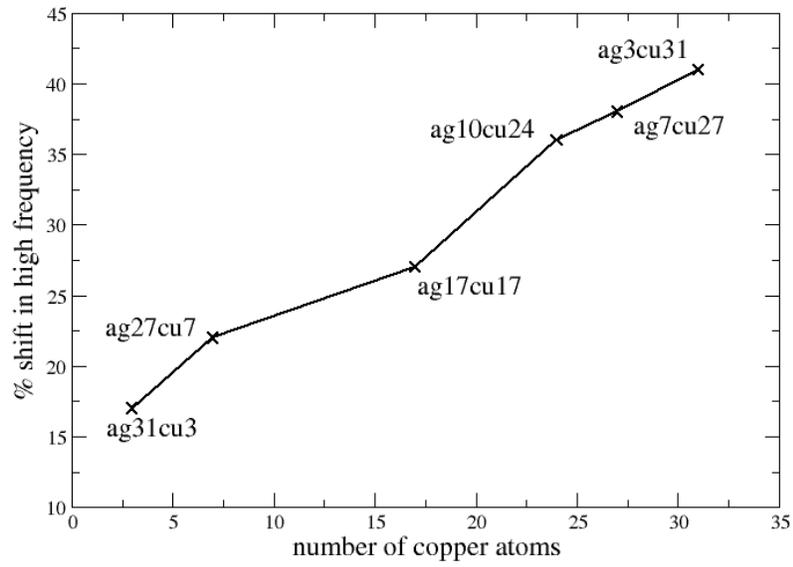

**Figure7.** Percentage shift at the high frequency end (above 5.1 THz) for chosen nano particles

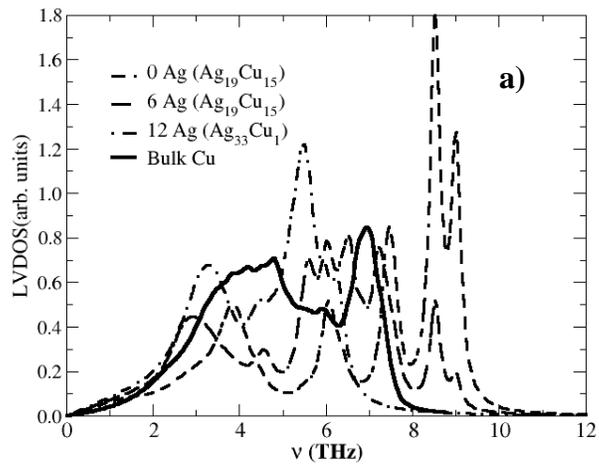

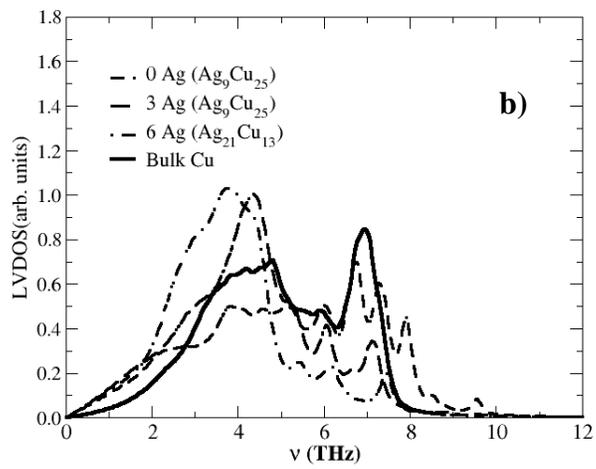

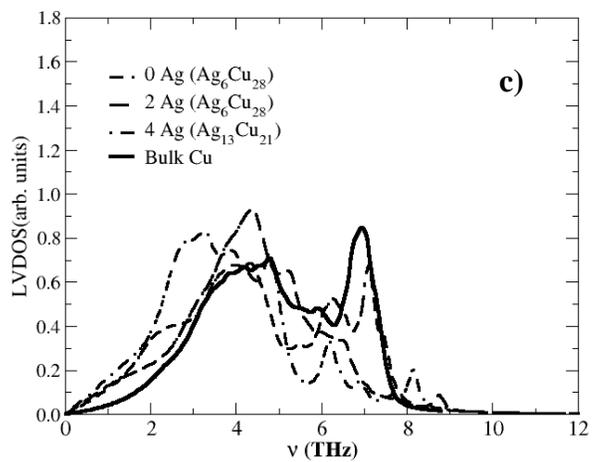

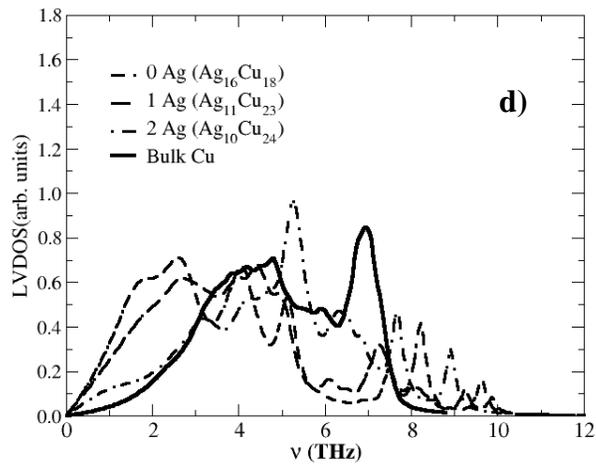

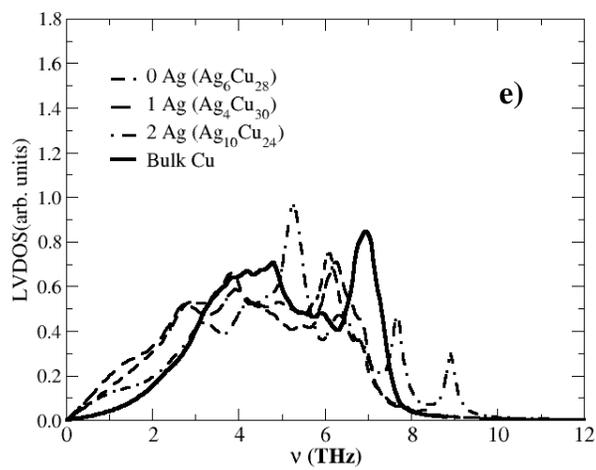

**Figure8.** Local Vibrational Densities of States (LVDOS) of chosen copper atoms with different coordination **a)** 12, **b)** 9, **c)** 8, and **d)** 6 (Large bond length) **e)** 6 (Short bond length) and elemental environment

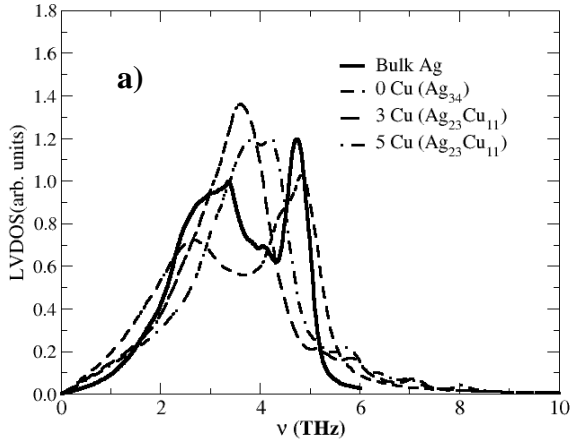

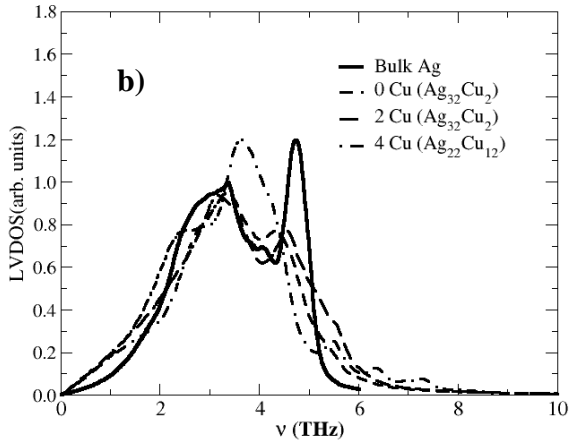

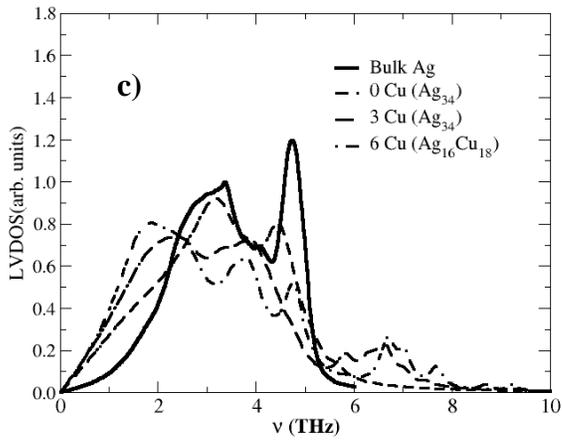

**Figure9.** Local Vibrational Densities of States (LVDOS) of chosen silver atoms with different coordination **a)** 9, **b)** 8, and **c)** 6 and elemental environment

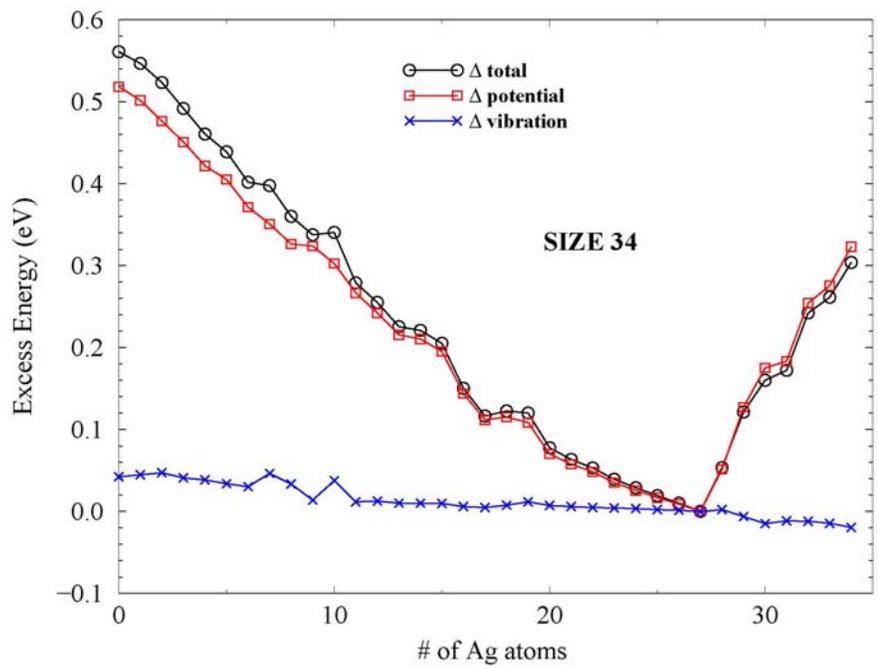

**Figure10.** The potential, vibrational and total excess energy of the nanoparticles

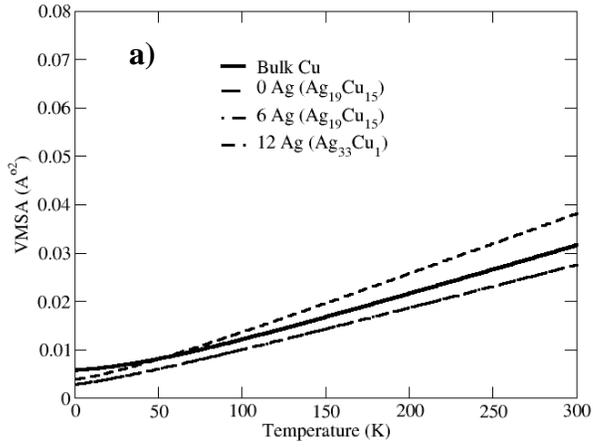

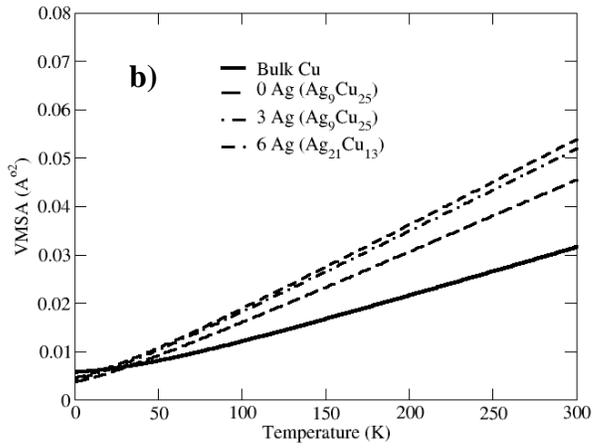

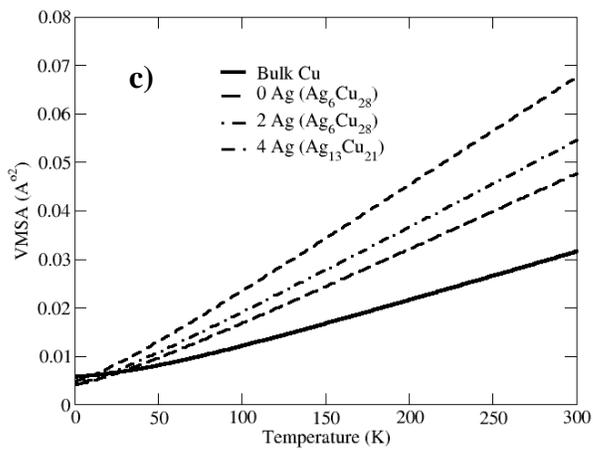

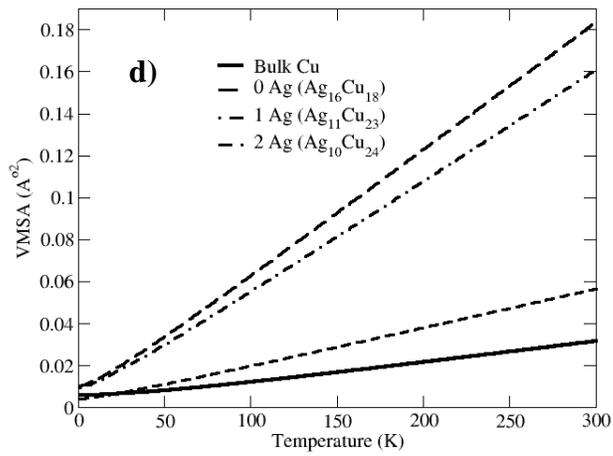

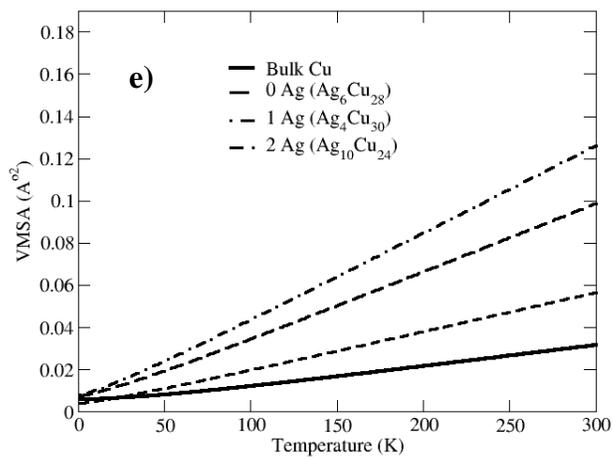

**Figure 11.** Mean Square Displacements of copper atoms with coordination **a)** 12, **b)** 9, **c)** 8, **d)** 6 (Large bond length) and **e)** 6 (Short bond length) and elemental environment

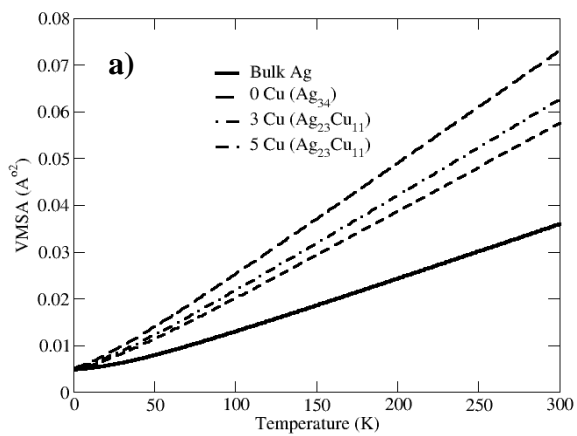

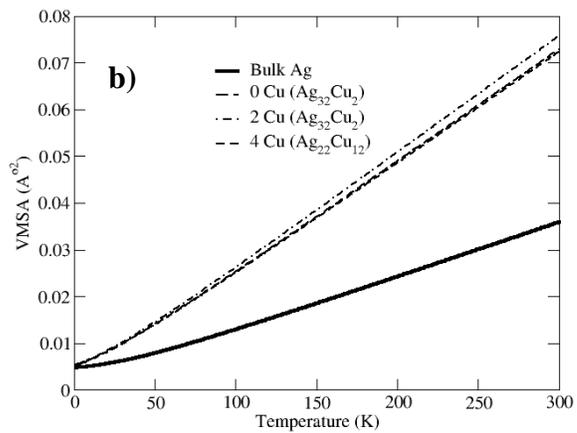

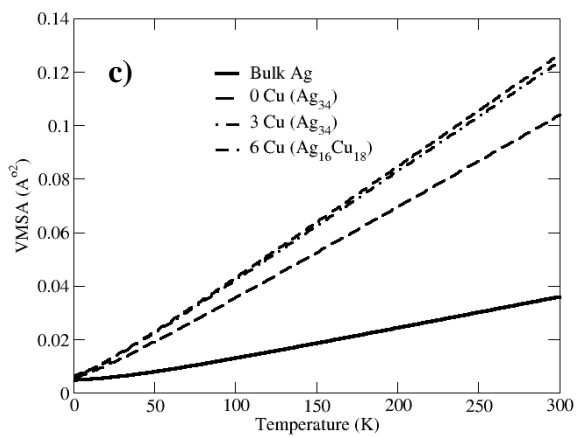

**Figure12.** Mean Square Displacements of silver atoms with coordination **a)** 9, **b)** 8 and **c)** 6 and elemental environment

**Table1.** Average bond length for Cu atom with coordination 6 along with elemental environment (DFT calculations)

| Cu coordination 6 | Number of Ag atoms and Bond Length (Å) | | |
|---|---|---|---|
| | 0Ag | 1Ag | 2Ag |
| $Ag_{18}Cu_{16}$ | 2.57(2.55) | - | - |
| $Ag_{16}Cu_{18}$ | 2.58(2.56) | - | - |
| $Ag_{15}Cu_{19}$ | 2.54(2.55) | - | - |
| $Ag_{14}Cu_{20}$ | 2.55(2.56) | - | - |
| $Ag_{13}Cu_{21}$ | 2.58(2.58) | - | - |
| $Ag_{12}Cu_{22}$ | 2.57(2.57) | - | - |
| $Ag_{11}Cu_{23}$ | 2.55(2.57) | 2.58(2.60) | - |
| $Ag_{10}Cu_{24}$ | 2.52(2.51) | 2.53(2.52) | 2.56(2.56) |
| $Ag_9Cu_{25}$ | 2.57(2.58) | - | - |
| $Ag_8Cu_{26}$ | 2.51(2.51) | 2.52(-) | - |
| $Ag_7Cu_{27}$ | 2.51(2.51) | 2.53(2.54) | - |
| $Ag_6Cu_{28}$ | 2.48(2.47) | 2.51(2.51) | - |
| $Ag_5Cu_{29}$ | 2.49(2.48) | 2.52(2.51) | - |
| $Ag_4Cu_{30}$ | 2.49(2.48) | 2.50(2.51) | - |
| $Ag_3Cu_{31}$ | 2.49(2.48) | 2.50(2.50) | - |
| $Ag_2Cu_{32}$ | 2.49(2.49) | 2.52(2.53) | - |
| $Ag_1Cu_{33}$ | 2.48(2.49) | 2.55(-) | - |
| Cu34 | 2.49(2.49) | - | - |

**Table2.** Total vibrational free energies of the nanoparticles and percentage contribution from low frequency end of the spectrum for some of the nanoparticles (using Debye frequency of bulk silver)

| Nanoparticle | $F_{vib}^{total}$ (eV) (% contribution from low frequency) |
|---|---|
| $Ag_{34}$ | -1.8226 |
| $Ag_{33}Cu_1$ | -1.7802 |
| $Ag_{32}Cu_2$ | -1.7529 |
| $Ag_{31}Cu_3$ | -1.7345 (73.4%) |
| $Ag_{30}Cu_4$ | -1.7378 |
| $Ag_{29}Cu_5$ | -1.6758 |
| $Ag_{28}Cu_6$ | -1.6167 |
| $Ag_{27}Cu_7$ | -1.6129 (71.4%) |
| $Ag_{26}Cu_8$ | -1.5906 |
| $Ag_{25}Cu_9$ | -1.5698 |
| $Ag_{24}Cu_{10}$ | -1.5492 |
| $Ag_{23}Cu_{11}$ | -1.5295 |
| $Ag_{22}Cu_{12}$ | -1.5117 |
| $Ag_{21}Cu_{13}$ | -1.4903 |
| $Ag_{20}Cu_{14}$ | -1.4673 |
| $Ag_{19}Cu_{15}$ | -1.4289 |
| $Ag_{18}Cu_{16}$ | -1.4361 |
| $Ag_{17}Cu_{17}$ | -1.4351(65.3%) |
| $Ag_{16}Cu_{18}$ | -1.4139 |
| $Ag_{15}Cu_{19}$ | -1.3773 |
| $Ag_{14}Cu_{20}$ | -1.3600 |
| $Ag_{13}Cu_{21}$ | -1.3457 |
| $Ag_{12}Cu_{22}$ | -1.3169 |
| $Ag_{11}Cu_{23}$ | -1.3060 |
| $Ag_{10}Cu_{24}$ | -1.1530 (53.4%) |
| $Ag_9Cu_{25}$ | -1.2652 |
| $Ag_8Cu_{26}$ | -1.1452 |
| $Ag_7Cu_{27}$ | -1.0826 (51.2%) |
| $Ag_6Cu_{28}$ | -1.1363 |
| $Ag_5Cu_{29}$ | -1.1018 |
| $Ag_4Cu_{30}$ | -1.0599 |
| $Ag_3Cu_{31}$ | -1.0322 (48.6%) |
| $Ag_2Cu_{32}$ | -0.9830 |
| $Ag_1Cu_{33}$ | -0.9812 |
| $Cu_{34}$ | -0.9795 |

**Table3.** The Average Debye temperature $\langle \theta_D \rangle$ of the nanoparticles, together with those from constituent Ag and Cu atoms

| Nanoparticle | $\langle \theta_D \rangle$ of Nanoparticles (K) | $\langle \theta_D \rangle$ of Cu Atoms (K) | $\langle \theta_D \rangle$ of Ag Atoms (K) |
|---|---|---|---|
| $Ag_{34}$ | 69.6 | - | 69.6 |
| $Ag_{33}Cu_1$ | 70.9 | 104.7 | 69.8 |
| $Ag_{32}Cu_2$ | 71.6 | 99.8 | 69.8 |
| $Ag_{31}Cu_3$ | 71.5 | 98.1 | 68.9 |
| $Ag_{30}Cu_4$ | 69.2 | 85.6 | 67.0 |
| $Ag_{29}Cu_5$ | 71.9 | 89.9 | 68.9 |
| $Ag_{28}Cu_6$ | 77.5 | 107.9 | 71.0 |
| $Ag_{27}Cu_7$ | 78.3 | 110.9 | 69.9 |
| $Ag_{26}Cu_8$ | 78.7 | 109.2 | 69.3 |
| $Ag_{25}Cu_9$ | 79.2 | 107.9 | 68.9 |
| $Ag_{24}Cu_{10}$ | 79.7 | 107.0 | 68.4 |
| $Ag_{23}Cu_{11}$ | 80.3 | 106.3 | 67.8 |
| $Ag_{22}Cu_{12}$ | 80.5 | 105.0 | 67.1 |
| $Ag_{21}Cu_{13}$ | 80.9 | 104.4 | 66.4 |
| $Ag_{20}Cu_{14}$ | 81.3 | 103.4 | 65.8 |
| $Ag_{19}Cu_{15}$ | 82.4 | 102.9 | 66.3 |
| $Ag_{18}Cu_{16}$ | 82.3 | 101.9 | 64.8 |
| $Ag_{17}Cu_{17}$ | 82.2 | 102.5 | 61.8 |
| $Ag_{16}Cu_{18}$ | 82.9 | 100.9 | 62.7 |
| $Ag_{15}Cu_{19}$ | 83.1 | 98.2 | 63.9 |
| $Ag_{14}Cu_{20}$ | 83.4 | 97.6 | 63.1 |
| $Ag_{13}Cu_{21}$ | 83.6 | 97.6 | 61.0 |
| $Ag_{12}Cu_{22}$ | 84.5 | 97.2 | 61.1 |
| $Ag_{11}Cu_{23}$ | 84.5 | 95.7 | 60.9 |
| $Ag_{10}Cu_{24}$ | 85.6 | 93.9 | 65.8 |
| $Ag_9Cu_{25}$ | 85.2 | 94.3 | 59.9 |
| $Ag_8Cu_{26}$ | 87.5 | 95.1 | 62.6 |
| $Ag_7Cu_{27}$ | 86.1 | 95.7 | 64.1 |
| $Ag_6Cu_{28}$ | 84.9 | 88.3 | 69.5 |
| $Ag_5Cu_{29}$ | 84.9 | 88.1 | 66.2 |
| $Ag_4Cu_{30}$ | 87.2 | 89.7 | 68.5 |
| $Ag_3Cu_{31}$ | 87.2 | 89.0 | 68.4 |
| $Ag_2Cu_{32}$ | 88.1 | 89.3 | 67.5 |
| $Ag_1Cu_{33}$ | 86.6 | 87.1 | 69.8 |
| $Cu_{34}$ | 87.4 | 87.4 | - |